\documentclass[preprint2]{emulateapj}

\newcommand{\bzcat}{ROMA-BZCAT}

\newcommand{\chn}{{\it Chandra}}

\newcommand{\fer}{{\it Fermi}}

\newcommand{\swf}{{\it Swift}}

\newcommand{\xmm}{{\it XMM-Newton}}
\newcommand{\wse}{{\it WISE}}

\usepackage{graphicx}
\usepackage{longtable}
\usepackage{lineno}

\slugcomment{version \today\ : fm}

\shorttitle{BL Lac candidates for TeV observations}
\shortauthors{F. Massaro et al. 2012}

\begin{document}
\title{BL Lac candidates for TeV observations}
\author{
F. Massaro\altaffilmark{1}, 
A. Paggi\altaffilmark{2}, 
M. Errando\altaffilmark{3}, 
R. D'Abrusco\altaffilmark{2},
N. Masetti\altaffilmark{4},
G. Tosti\altaffilmark{5,6}, 
\& 
S. Funk\altaffilmark{1}
}

\altaffiltext{1}{SLAC National Laboratory and Kavli Institute for Particle Astrophysics and Cosmology, 2575 Sand Hill Road, Menlo Park, CA 94025, USA}
\altaffiltext{2}{Smithsonian Astrophysical Observatory, 60 Garden Street, Cambridge, MA 02138, USA}
\altaffiltext{3}{Department of Physics and Astronomy, Barnard College, Columbia University, New York, NY 10027, USA}
\altaffiltext{4}{INAF - Istituto di Astrofisica Spaziale e Fisica Cosmica di Bologna, via Gobetti 101, 40129, Bologna, Italy}
\altaffiltext{5}{Dipartimento di Fisica, Universit\`a degli Studi di Perugia, 06123 Perugia, Italy}

\begin{abstract} 
BL Lac objects are the most numerous class of extragalactic TeV-detected sources.
One of the biggest difficulties in investigating their TeV emission resides in their limited number, 
since only 47 BL Lacs are known as TeV emitters.
In this paper, we propose new criteria to select TeV BL Lac candidates based on the infrared (IR) and X-ray observations. 
We apply our selection criteria to the BL Lac objects listed in the \bzcat\ catalog so identifying 41 potential TeV emitters.
We then consider a search over a more extended sample combining the ROSAT
bright source catalog and the \wse\ all-sky survey revealing 54 additional candidates for TeV observations.
Our investigation also led to a tentative classification of 16 unidentified X-ray sources as BL Lac candidates.
{ This analysis provides new interesting BL Lac targets for future observations with ground based Cherenkov telescopes.}
\end{abstract}


\section{Introduction}
BL Lac objects are characterized by very peculiar properties with respect to other classes of active galactic nuclei (AGNs).
They are compact, core dominated radio sources, many of them exhibiting superluminal motion and showing
rapid and large-amplitude flux variability from radio up to TeV energies, 
and significant radio to optical polarization \citep[e.g.,][]{blandford78,urry95}.
Their spectral energy distribution (SED) exhibits two main components: the low energy one peaking 
in the infrared-X-ray energy range, and the second one dominated by $\gamma$-rays.
Their optical spectra appear to be featureless or with very weak absorption lines \citep{stoke91,stickel91,laurent99}.

According to Padovani \& Giommi (1995) BL Lacs can be classified 
as ``Low-frequency peaked BL Lacs'' (LBLs) and ``High-frequency peaked BL Lacs'' (HBLs), 
depending on whether their broadband radio-to-X-ray spectral index 
is larger than or smaller than 0.75, respectively. 

At very high energies (i.e., $E>$100 GeV) BL Lac objects, and in particular, 
HBLs, constitute the largest known population of TeV extragalactic sources,
detected by ground based Cherenkov telescopes as HESS, MAGIC and VERITAS.
In the following, we refer to the HBLs detected at TeV energies as TBLs
while we indicate the HBL candidates for TeV observations as TBCs.

{ Recently, using the \wse\ point source catalog, 
which mapped the sky in four different bands centered at 3.4, 4.6, 12, and 22 $\mu$m \citep{wright10,cutri12}, 
we discovered that $\gamma$-ray emitting blazars occupy a distinct region in the two-dimensional color-color diagrams, 
which is well separated from other extragalactic sources whose IR emission is dominated by thermal radiation 
\citep[``the \wse\ Gamma-ray Strip'',][]{paper1,paper2} and we have developed a method for identifying 
$\gamma$-ray blazar candidates by studying the \wse\ three-dimensional IR color space 
using the \wse\ Fermi Blazar Sample  \citep[i.e., ``locus'', see][]{paper6}. This discovery
constitutes the basis of our selection criterion for the TBCs.}

In this paper, we combine IR and X-ray archival data available in literature
to build a criterion useful to find new TBCs.
We use the X-ray observations performed with ROSAT along with those from the
Wide Infrared Survey Explorer (\wse) satellite \citep{wright10}.

This paper is organized as follows: in \S~\ref{sec:tev} we investigate the IR properties of the BL Lacs
already detected at TeV energies introducing the ``$\Phi_{XIR}$ parameter" to distinguish between LBLs and HBLs.
In \S~\ref{sec:criteria} we outline our criterion to identify TBCs and apply it to the 
{ BL Lacs listed in the \bzcat\footnote{http://www.asdc.asi.it/bzcat/} \citep[e.g.,][]{massaro11} and
comparisons with selection criteria previously published are presented in \S~\ref{sec:comparison}.}
\S~\ref{sec:rasswise} is dedicated to the all-sky search of new TBL candidates 
using the combination of the ROSAT bright source catalog \citep{voges99} and the \wse\ all-sky survey \citep{wright10} and
\S~\ref{sec:summary} is devoted to our summary and conclusions.
  
\wse\ magnitudes are in the Vega system
and we use cgs units for our numerical results  unless stated otherwise. 
We assume a flat cosmology with $H_0=72$ km s$^{-1}$ Mpc$^{-1}$,
$\Omega_{M}=0.26$ and $\Omega_{\Lambda}=0.74$ (Dunkley et al. 2009).
Spectral indices, $\alpha$, are defined by flux density, S$_{\nu}\propto\nu^{-\alpha}$.
Frequent acronyms are listed in Table~\ref{tab:acronym}.
\begin{table}
\begin{center}
\caption{List of acronyms.}
\begin{tabular}{|lc|}
\hline
Name & Acronym \\
\hline
\noalign{\smallskip}
High Frequency Peaked BL Lac & HBL\ \\ 
Low Frequency Peaked BL Lac & LBL \\
HBL detected at TeV energies & TBL \\
HBL candidate for TeV observations & TBC \\
\noalign{\smallskip}
\hline
\end{tabular}\\
\label{tab:acronym}
\end{center}
\end{table}

\section{TeV BL Lac objects}
\label{sec:tev}
{ According to the online catalog of TeV-emitting $\gamma$-ray sources (TeVCat)}\footnote{http://tevcat.uchicago.edu/}, 
the number of sources classified as BL Lac objects in the \bzcat\ \citep{massaro09,massaro10,massaro11} 
and detected at TeV energies is 42, as of December 2012;
these TeV BL Lac objects are listed in Table~\ref{tab:main} together with their salient parameters.
They have a unique \wse\ counterpart detected at least at 3.4, 4.6 and 12 $\mu$m
within a radius of 3\arcsec.3 from the \bzcat\ positions \citep[see][for more details about the \bzcat\ - \wse\ positional associations]{paper6}.
They also have a radio counterpart and are detected in the X-ray band by 
ROSAT \citep{voges99} as reported in the \bzcat\ \citep{massaro11} with the only exception of MAGICJ2001+435.
Thirty-seven of them are also detected in $\gamma$ rays between 30 MeV and 100 GeV  
as reported in the \fer-LAT second source Catalog \citep[2FGL;][]{nolan12} and in the second \fer-LAT 
AGN catalog \citep[2LAC;][]{ackermann11}.

\subsection{X-ray-to-infrared flux ratio: $\Phi_{XIR}$}
\label{sec:phi}
Maselli et al. (2010a) defined the ratio  $\Phi_{XR}$ between the ROSAT X-ray flux $F_X$ 
and the radio flux density $S_{1.4}$ (at 1.4 GHz), computed using the values reported in the \bzcat,
to distinguish between HBLs  (i.e., $\Phi_{XR}$ $\geq$ 0.1) and LBLs (i.e., $\Phi_{XR}<$ 0.1). 
However, this distinction cannot be easily extended all-sky because it does need 
radio observations at 1.4 GHz which are not always available.
To avoid this problem we define a new parameter to distinguish 
between the two subclasses of BL Lac objects based on the IR observations of \wse.
It is worth noting that among the BL Lac objects the HBLs are the most detected at TeV energies.

For all the TeV BL Lac objects listed in Table~\ref{tab:main},
we computed $\Phi_{XIR}$, defined as the ratio between the ROSAT X-ray flux $F_X$ (0.1 - 2.4 keV) 
and the integrated IR flux $F_{IR}$ between 3.4 and 12 $\mu$m,
both in units of 10$^{-12}$ erg\,cm$^{-2}$\,s$^{-1}$.
This parameter is used to distinguish between HBLs and LBLs instead of $\Phi_{XR}$.
We note that sources with $\Phi_{XIR}>$0.1 in Table~\ref{tab:main} have a $\Phi_{XR}>$0.08 
in agreement with the previous classification \citep{maselli10a,maselli10b}.
We then consider a new classification, labeling as
HBLs those having $\Phi_{XIR}>$0.1 while indicating as LBLs those with $\Phi_{XIR}<$0.1.
An additional justification on the choice of $\Phi_{XIR}$ to classify BL Lacs is given in Appendix
on the basis of their spectral shape.

\subsection{TBL sample selection}
\label{sec:tbl}
We define a clean sample of 33 HBLs TeV detected (i.e., TBLs) out of 42 TeV sources including only:
\begin{itemize}
\item having $\Phi_{XIR}>$0.1;
\item with \wse\ magnitudes lower than 13.32 mag, 12.64 mag and 10.76 mag 
at 3.4$\mu$m, 4.6$\mu$m and 12$\mu$m, respectively;
\item with IR colors between 0.22 mag$<$[3.4]-[4.6]$<$0.86 mag, 1.60 mag$<$[4.6]-[12]$<$2.32 mag.
\end{itemize}
The above criterion of TBLs, not only based on the $\Phi_{XR}$ ratio,
{ permits} to select bright IR sources having the first SED peak between the UV and the X-rays.
{ The minimum X-ray and IR fluxes of the resulting sample are}
2.45 $\cdot$ 10$^{-12}$ erg\,cm$^{-2}$\,s$^{-1}$ and 9.47 $\cdot$ 10$^{-13}$ erg\,cm$^{-2}$\,s$^{-1}$, respectively.
\begin{figure}[!ht]
\includegraphics[height=8.7cm,width=6.8cm,angle=-90]{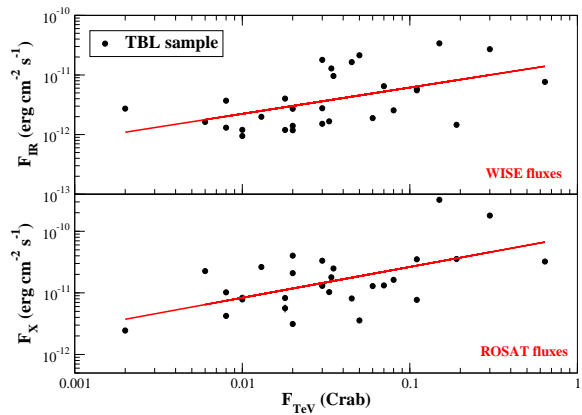}
\caption{The correlations between the IR (upper panel) and the X-ray (lower panel) fluxes with the TeV flux,
reported in the \wse\ ROSAT and TeVCat catalogs, respectively (see also Table~\ref{tab:main}).
Regression lines are shown in red (see Section~\ref{sec:tbl} for more details).}
\label{fig:flux}
\end{figure}

As shown in Figure~\ref{fig:flux}, there is a hint of a correlation between the IR and TeV fluxes 
for the TBLs whose measurements were available in TeVCat (see also Table~\ref{tab:main}), 
with a correlation coefficient of 0.51.
This suggests a good match between \wse\ and the TeV observations.
As expected there is also a trend between the ROSAT and the TeV fluxes,
with a correlation coefficient of 0.58.

\subsection{Infrared colors of TBLs}
\label{sec:infrared}
\begin{figure*}[]
\includegraphics[height=9.cm,width=7.cm,angle=-90]{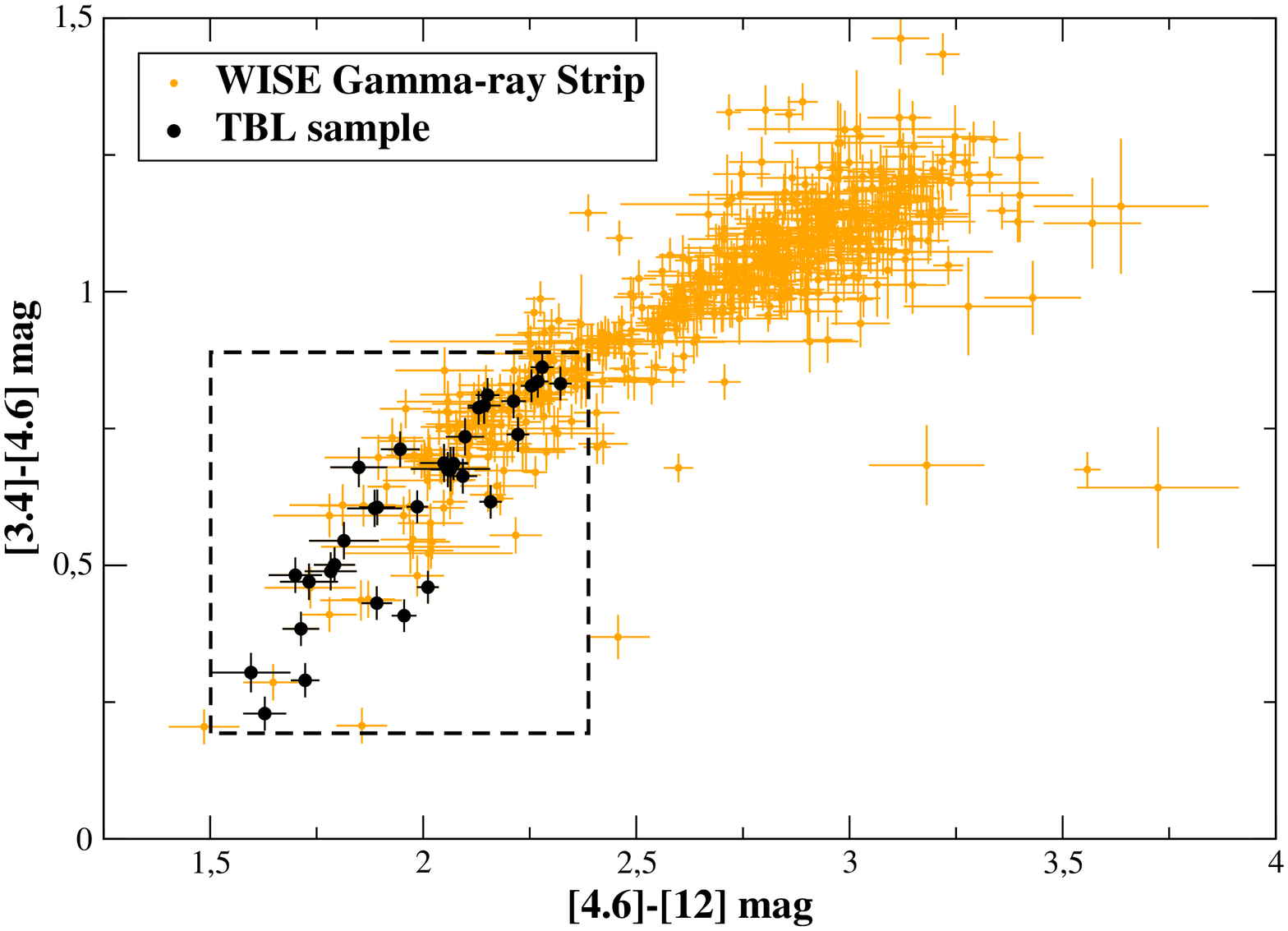}
\includegraphics[height=9.cm,width=7.cm,angle=-90]{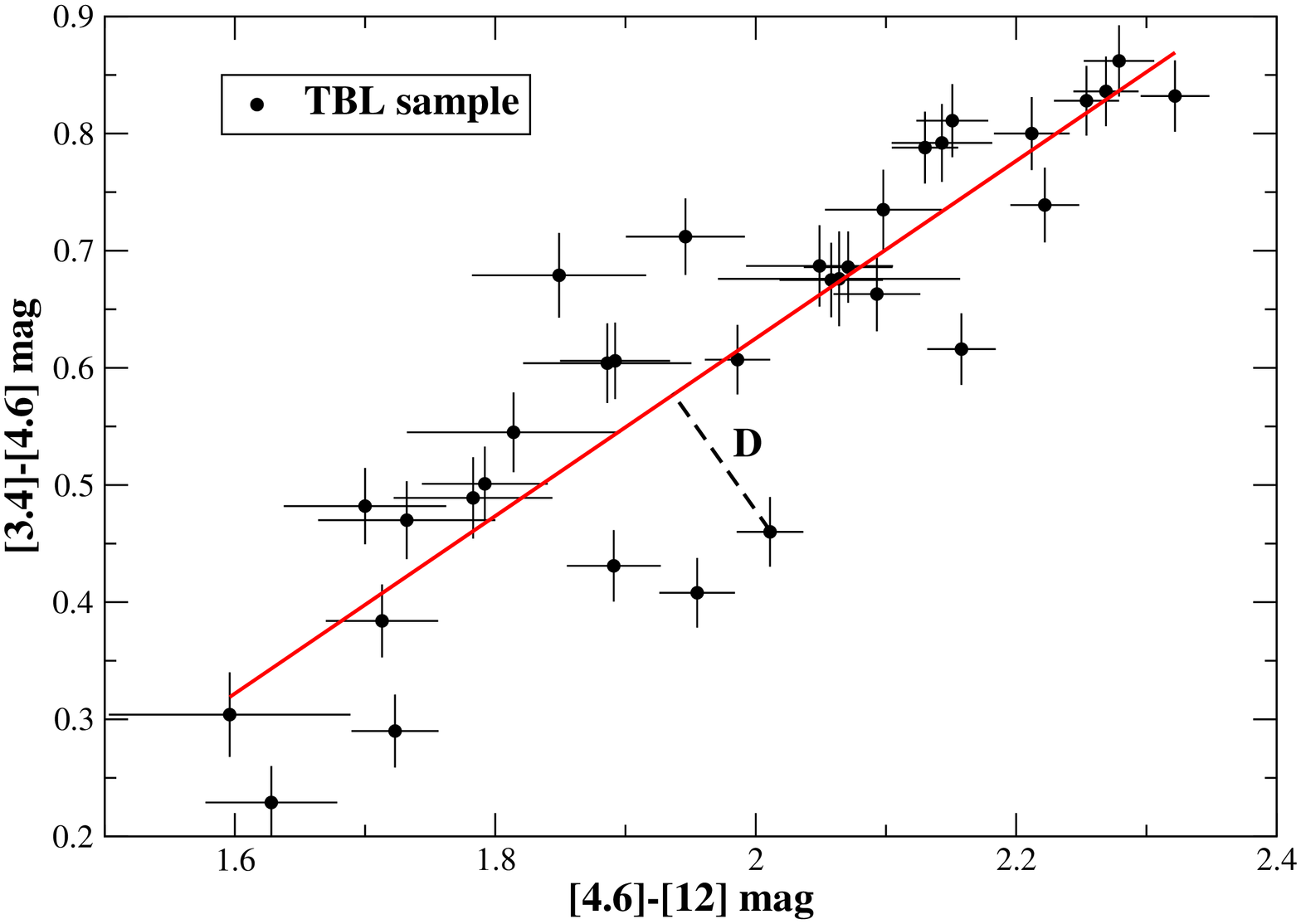}
\caption{Left panel: the [3.4]-[4.6]-[12] $\mu$m color-color plot for the 33 TBLs selected (black circles)
overlaid to the $\gamma$-ray emitting blazars associated with \wse\ source that constitute the 
\wse\ Gamma-ray Strip \citep[see][for more details]{paper1,paper2,paper6}. The black dashed box
indicates the subregion of the \wse\ Gamma-ray Strip considered in our TBC selection.
Right panel: the [3.4]-[4.6]-[12] $\mu$m color-color plot for the 33 TBLs selected (black circles).
The red line corresponds to the regression line evaluated while
the dashed line indicates the distance $D$ between a source and the regression line as described in \S~\ref{sec:infrared}.}
\label{fig:strip}
\end{figure*}

Recently, we discovered that the $\gamma$-ray BL Lac objects 
lie in a region (i.e., the \wse\ Gamma-ray Strip) 
of the [3.4]-[4.6]-[12] $\mu$m color-color diagram well differentiated 
from that occupied by generic IR sources \citep{paper1}. 
In particular, TBLs are more concentrated near the {\it tail} of the \wse\ Gamma-ray Strip.
In Figure~\ref{fig:strip} we show the IR colors of TBLs and those of 
$\gamma$-ray BL Lacs detected by \fer\ in the 2LAC CLEAN 
sample that belong to the \wse\ Gamma-ray Strip \citep{paper6}.

We calculated a linear regression in the [3.4]-[4.6]-[12] $\mu$m color-color plot
for the 33 selected TBLs as shown in Figure~\ref{fig:strip}.
We then define the $\delta$ parameter according to the equation:
\begin{equation}
\delta = \left| D \cdot D_{max}^{-1} \right|
\end{equation}
where $D$ is the distance between the IR colors of 
each source in the [3.4]-[4.6]-[12] $\mu$m color-color diagram and the regression line, 
and $D_{max}$ is the maximum value evaluated only for the selected TBLs (i.e., 0.116 mag).
{ All the TBLs have therefore a value of delta < 1.0 by definition.}
We verified the residuals of the points in the [3.4]-[4.6]-[12] $\mu$m color-color plot
with respect to the regression line using a {\it runs test} and we found that they are randomly distributed 
at 97\% level of confidence
\footnote{{ The {\it runs test} is a non-parametric statistical test that verifies the hypothesis that the elements 
of the sequence are mutually independent and it could be applied in combination with a regression analysis
to check that residuals are randomly distributed as expected in a Gaussian statistic.}}.

Finally, we note that $\delta$=0.28 corresponds to the 68\% level of confidence (i.e., 1$\sigma$) with respect to the regression line
and consequently the choice of $\delta$=1 implies that all the TBLs lie within 3.5$\sigma$.

\section{TBL candidates selected from the \bzcat\ catalog}
\label{sec:criteria}
On the basis of the combined IR and X-ray properties of TBLs (see \S~\ref{sec:tev}),
we outline the following criteria to select TBL candidates (TBCs). 
Our selection of TBCs includes all sources that fulfill all following criteria:
\begin{enumerate}
\item classified as BL Lac (i.e., BZB) according to the \bzcat\ catalog;
\item have a \wse\ counterpart within 3.3\arcsec\ from the \bzcat\ position, detected with Vega magnitudes
smaller than 13.318, 12.642 and 10.760 at 3.4$\mu$m, 4.6$\mu$m and 12$\mu$m, respectively;
\item have IR colors similar to those of TBLs defined by: 0.22 mag$<$[3.4]-[4.6]$<$0.86 mag, 1.60 mag$<$[4.6]-[12]$<$2.32 mag, respectively;
\item have values of the parameter $\delta<$1, according to the definition proposed in \S~\ref{sec:infrared};
\item have X-ray fluxes larger than the minimum value observed for TBLs (i.e., 2.45 $\cdot$ 10$^{-12}$ erg\,cm$^{-2}$\,s$^{-1}$).
\end{enumerate} 
The requirement on the X-ray and IR fluxes ensures that the selected sources will be above ROSAT and \wse\ sensitivity thresholds.

{ We apply our selection to the BL Lac objects 
that belong to the \bzcat\ \citep{massaro09,massaro10,massaro11}.}

For the blazars listed in the \bzcat\ we found 41 TBCs that meet our criteria.
All these TBCs are detected by \fer\ in the 30 MeV - 300 GeV energy range 
with only five exceptions: BZB J0056-0936, BZB J0214+5144, BZB J0809+3455, BZB J1215+0732 and BZB J1445-0326,
in particular, about $\sim$90\% of them show hard $\gamma$-ray spectra (i.e., $\gamma$-ray photon index $\Gamma<$2).
Their complete list can be found in Table~\ref{tab:candidates}, where we report their \bzcat\ name, that of their \wse\ counterpart,
the redshift if known, the ROSAT X-ray flux corrected for the Galactic absorption \citep{kalberla05}, 
the \fer\ $\gamma$-ray spectral index, the IR \wse\ colors together with the IR flux in the 3.4-12$\mu$m
band and the value of $\Phi_{XIR}$.

Finally, we note that , all the TBCs selected from the \bzcat\ lie within 3$\sigma$ level of confidence of the regression line (see Section~\ref{sec:infrared}),
with the only exceptions of  BZB J0214+5144 and BZB J1445-0326.

\begin{table*}
\tiny
\caption{The complete list of TeV detected BL Lac objects (00 -- 24 HH).}
\begin{tabular}{|lllcccccccc|}
\hline
\bzcat\ & TeVCat  & \wse\ & z & F$_X$ & F$_{TeV}$ & \fer\ & [3.4]-[4.6] & [4.6]-[12] & F$_{IR}$ & $\Phi_{XIR}$\\
  name  &  name   &  name &  & cgs & Crab & detect. & mag & mag & cgs & \\
\hline
\noalign{\smallskip}
J0013-1854 & SHBL\,J001355.9-185406 & J001356.04-185406.5 & 0.094  & 6.49  & 0.01   & no    & 0.22(0.03) & 1.32(0.10)     & 1.36(0.08) & 4.76\\ 
J0033-1921 & KUV\,00311-1938        & J003334.36-192132.9 & 0.61?   & 8.43  & ---    & yes   & 0.79(0.03) & 2.14(0.04)     & 3.06(0.08) & 2.75 \\    
J0035+5950 & 1ES\,0033+595          & J003552.62+595004.3 & ?      & 5.41  & ---    & yes   & 0.66(0.03) & 2.09(0.03)     & 2.46(0.06) & 2.20 \\    
J0152+0147 & RGB\,J0152+017         & J015239.60+014717.4 & 0.08   & 3.13  & 0.02   & yes   & 0.38(0.03) & 1.71(0.04)     & 2.71(0.08) & 1.16\\    
J0222+4302 & 3C\,66A                & J022239.60+430207.8 & 0.444? & 2.29  & 0.022  & yes   & 0.84(0.03) & 2.25(0.03)     & 36.87(0.70) & 0.06\\    
J0232+2017 & 1ES\,0229+200          & J023248.60+201717.3 & 0.139  & 5.63  & 0.018  & no    & 0.30(0.04) & 1.60(0.09)     & 1.19(0.07) & 4.71\\    
J0303-2407 & PKS\,0301-243          & J030326.49-240711.4 & 0.26?  & 5.78  & ---    & yes   & 0.86(0.03) & 2.28(0.03)     & 8.50(0.17) & 0.68\\    
J0319+1845 & RBS\,0413              & J031951.80+184534.6 & 0.19   & 8.33  & 0.01   & yes   & 0.68(0.04) & 2.06(0.09)     & 0.95(0.06) & 8.79\\    
J0349-1159 & 1ES\,0347-121          & J034923.18-115927.2 & 0.188  & 14.28 & 0.02   & no    & 0.51(0.05) & 1.57(0.27)     & 0.33(0.06) & 43.57\\    
J0416+0105 & 1ES\,0414+009          & J041652.48+010523.9 & 0.287  & 22.6  & 0.006  & yes   & 0.68(0.04) & 1.85(0.07)     & 1.63(0.07) & 13.86\\    
J0449-4350 & PKS\,0447-439          & J044924.69-435008.9 & 0.205? & 8.12  & 0.045  & yes   & 0.84(0.03) & 2.27(0.02)     & 16.43(0.31) & 0.49\\    
J0507+6737 & 1ES\,0502+675          & J050756.16+673724.3 & 0.416? & 12.9  & 0.06   & yes   & 0.71(0.03) & 1.95(0.05)     & 1.89(0.06) & 6.82\\    
J0550-3216 & PKS\,0548-322          & J055040.57-321616.4 & 0.069  & 26.3  & 0.013  & no    & 0.23(0.03) & 1.63(0.05)     & 1.99(0.06) & 13.24\\    
J0648+1516 & RX\,J0648.7+1516       & J064847.64+151624.8 & 0.179  & 10.3  & 0.033  & yes   & 0.60(0.03) & 1.89(0.06)     & 1.67(0.07) & 6.18\\    
J0650+2503 & 1ES\,0647+250          & J065046.48+250259.6 & 0.203? & 13.0  & 0.03   & yes   & 0.73(0.03) & 2.10(0.04)     & 2.76(0.09) & 4.70\\    
J0710+5908 & RGB\,J0710+591         & J071030.05+590820.5 & 0.125  & 13.4  & 0.03   & yes   & 0.49(0.03) & 1.78(0.06)     & 1.52(0.06) & 8.82\\    
J0721+7120 & S5\,0716+714           & J072153.44+712036.3 & ?      & 2.27  & ---    & yes   & 0.98(0.03) & 2.66(0.02)     & 69.98(1.29) & 0.03\\    
J0809+5218 & 1ES\,0806+524          & J080949.19+521858.3 & 0.138  & 8.26  & 0.018  & yes   & 0.69(0.03) & 2.07(0.03)     & 4.02(0.10) & 2.06\\    
J1010-3119 & 1RXS\,J101015.9-311909 & J101015.98-311908.3 & 0.143  & 10.2  & 0.008  & yes   & 0.47(0.03) & 1.73(0.07)     & 1.31(0.06) & 7.81\\    
J1015+4926 & 1ES\,1011+496          & J101504.13+492600.8 & 0.212  & 13.2  & 0.07   & yes   & 0.80(0.03) & 2.21(0.03)     & 6.49(0.14) & 2.03\\    
J1103-2329 & 1ES\,1101-232          & J110337.62-232931.0 & 0.186  & 20.9  & 0.02   & yes   & 0.54(0.03) & 1.81(0.08)     & 1.18(0.06) & 17.68\\    
J1104+3812 & Markarian\,421         & J110427.32+381231.9 & 0.03   & 180.0 & 0.3    & yes   & 0.61(0.03) & 1.99(0.02)     & 27.08(0.50) & 6.65\\    
J1136+7009 & Markarian\,180         & J113626.42+700927.1 & 0.045  & 35.1  & 0.11   & yes   & 0.41(0.03) & 1.96(0.03)     & 5.87(0.12) & 5.98\\    
J1217+3007 & 1ES\,1215+303          & J121752.08+300700.7 & 0.13?  & 24.9  & 0.035  & yes   & 0.83(0.03) & 2.32(0.03)     & 9.62(0.19) & 2.59\\    
J1221+3010 & 1ES\,1218+304          & J122121.95+301037.2 & 0.182  & 16.3  & 0.08   & yes   & 0.68(0.03) & 2.06(0.04)     & 2.55(0.07) & 0.10\\    
J1221+2813 & W\,Comae               & J122131.69+281358.5 & 0.102  & 1.3   & 0.09   & yes   & 0.85(0.03) & 2.33(0.03)     & 13.5(0.27) & 6.40\\    
J1315-4236 & 1ES\,1312-423          & J131503.39-423649.7 & 0.105  & 8.85  & 0.004  & no    & 0.28(0.04) & 1.23(0.16)     & 0.66(0.06) & 13.48\\    
J1427+2348 & PKS\,1424+240          & J142700.40+234800.1 & ?      & 3.57  & 0.05   & yes   & 0.83(0.03) & 2.25(0.02)     & 21.43(0.40) & 0.17\\    
J1428+4240 & H\,1426+428            & J142832.62+424021.0 & 0.129  & 35.5  & 0.19   & yes   & 0.50(0.03) & 1.79(0.05)     & 1.46(0.05) & 24.33\\    
J1442+1200 & 1ES\,1440+122          & J144248.24+120040.3 & 0.163  & 7.82  & 0.01   & yes   & 0.48(0.03) & 1.70(0.06)     & 1.20(0.05) & 6.50\\    
J1517-2422 & AP\,Lib                & J151741.82-242219.4 & 0.048  & 1.05  & 0.02   & yes   & 0.88(0.03) & 2.61(0.03)     & 27.94(0.53) & 0.04\\    
J1555+1111 & PG\,1553+113           & J155543.05+111124.4 & ?      & 17.9  & 0.034  & yes   & 0.81(0.03) & 2.15(0.03)     & 12.86(0.26) & 1.39\\    
J1653+3945 & Markarian\,501         & J165352.22+394536.5 & 0.033  & 36.9  & ---    & yes   & 0.46(0.03) & 2.01(0.03)     & 19.5(0.37) & 1.89\\    
J1743+1935 & 1ES\,1741+196          & J174357.84+193509.3 & 0.084  & 4.23  & 0.008  & yes   & 0.43(0.03) & 1.89(0.04)     & 3.70(0.09) & 1.14\\    
J1959+6508 & 1ES\,1959+650          & J195959.84+650854.7 & 0.047  & 32.3  & 0.64   & yes   & 0.62(0.03) & 2.16(0.03)     & 7.66(0.15) & 4.22\\    
J2001+4352 & MAGIC\,J2001+435       & J200112.87+435252.8 & ?      & ---   & 0.22   & yes   & 0.77(0.03) & 2.16(0.03)     & 7.56(0.16) & ---\\    
J2009-4849 & PKS\,2005-489          & J200925.39-484953.5 & 0.071  & 33.3  & 0.03   & yes   & 0.74(0.03) & 2.22(0.03)     & 17.92(0.35) & 1.86\\    
J2158-3013 & PKS\,2155-304          & J215852.05-301332.0 & 0.116  & 324.0 & 0.15   & yes   & 0.79(0.03) & 2.13(0.03)     & 33.93(0.65) & 9.55\\    
J2202+4216 & BL\,Lacertae           & J220243.29+421640.0 & 0.069  & 1.58  & 0.03   & yes   & 1.01(0.03) & 2.60(0.03)     & 126.47(2.41) & 0.01\\    
J2250+3824 & B3\,2247+381           & J225005.75+382437.3 & 0.119  & 2.45  & 0.002  & yes   & 0.61(0.03) & 1.89(0.04)     & 2.73(0.08) & 0.90\\    
J2347+5142 & 1ES\,2344+514          & J234704.83+514217.9 & 0.044  & 7.71  & 0.11   & yes   & 0.29(0.03) & 1.72(0.03)     & 5.52(0.13) & 1.40\\    
J2359-3037 & H\,2356-309            & J235907.88-303740.5 & 0.165  & 40.2  & 0.02   & yes   & 0.69(0.03) & 2.05(0.06)     & 1.40(0.05) & 28.67\\    
\noalign{\smallskip}
\hline
\end{tabular}\\
Col. (1) \bzcat\ name. \\
Col. (2) TeVCat name.\\
Col. (3) \wse\ name.  \\
Col. (4) \bzcat\ redshift: ? = unknown, number? = uncertain.  \\
Col. (5) ROSAT X-ray flux in the 0.1-2.4 keV energy range in units of 10$^{-12}$ erg\,cm$^{-2}$\,s$^{-1}$. \\
Col. (6) archival TeV flux as reported on the TeVCat. \\
Col. (7) \fer\ detection as reported in the 2FGL. \\
Cols. (8,9) IR colors from \wse. Values in parentheses are 1$\sigma$ uncertainties. \\
Col. (10) \wse\ IR flux in the 3.4-12$\mu$m energy range in units of 10$^{-12}$ erg\,cm$^{-2}$\,s$^{-1}$.  \\
Col. (11) $\Phi_{XIR}$ defined according to \S~\ref{sec:tev}. 
\label{tab:main}
\end{table*}

\begin{table*}
\tiny
\caption{The complete list of TBCs selected from the \bzcat\ (00 -- 24 HH).}
\begin{tabular}{|llcccccccr|}
\hline
\bzcat\ & \wse\ & z & F$_X$ & $\Gamma$ &[3.4]-[4.6] & [4.6]-[12] & F$_{IR}$ & $\Phi_{XIR}$  & Sel.\\
  name  &  name &   & cgs   &          & mag        & mag        & cgs      &          & \\
\hline
\noalign{\smallskip}
  BZB J0035+1515  &  J003514.71+151504.2  &  ?      &  3.02  &  1.62  &  0.78(0.03)  &  2.20(0.06)  &  1.63(0.07)  &  1.86 & --- \\
  BZB J0056-0936  &  J005620.06-093630.6  &  0.103  &  4.20  &  ---   &  0.29(0.03)  &  1.75(0.05)  &  2.52(0.08)  &  1.67 & --- \\
  BZB J0109+1816  &  J010908.17+181607.7  &  0.145  &  4.51  &  1.99  &  0.81(0.03)  &  2.28(0.05)  &  1.87(0.06)  &  2.42 & T \\ 
  BZB J0136+3905$^*$  &  J013632.59+390559.2  &  ?      &  9.60  &  1.69  &  0.79(0.03)  &  2.11(0.03)  &  4.01(0.09)  &  2.39 & C,T \\ 
  BZB J0209-5229  &  J020921.60-522922.7  &  ?      &  7.98  &  1.91  &  0.61(0.03)  &  1.97(0.05)  &  1.58(0.05)  &  5.06 & --- \\ 
  BZB J0214+5144  &  J021417.94+514451.9  &  0.049  &  4.58  &  ---   &  0.29(0.03)  &  1.77(0.04)  &  3.48(0.09)  &  1.31 & M \\ 
  BZB J0238-3116  &  J023832.47-311657.9  &  ?      &  5.14  &  1.85  &  0.64(0.03)  &  1.91(0.04)  &  1.96(0.06)  &  2.62 & --- \\
  BZB J0316-2607  &  J031614.93-260757.2  &  0.443  &  3.05  &  1.87  &  0.75(0.03)  &  2.14(0.04)  &  1.34(0.04)  &  2.27 & --- \\
  BZB J0325-1646  &  J032541.09-164616.8  &  0.291  &  27.2  &  1.97  &  0.70(0.04)  &  2.02(0.07)  &  1.10(0.05)  & 24.77 & --- \\ 
  BZB J0326+0225  &  J032613.94+022514.7  &  0.147  &  12.0  &  2.06  &  0.58(0.04)  &  2.02(0.08)  &  1.19(0.06)  & 10.07 & C,M,S \\ 
  BZB J0505+0415  &  J050534.76+041554.5  &  0.027? &  3.07  &  2.15  &  0.60(0.04)  &  2.03(0.08)  &  1.02(0.06)  &  3.02 & --- \\
  BZB J0536-3343  &  J053629.06-334302.5  &  ?      &  4.84  &  2.39  &  0.68(0.03)  &  2.04(0.05)  &  1.31(0.05)  &  3.70 & --- \\
  BZB J0543-5532  &  J054357.21-553207.5  &  ?      &  9.04  &  1.74  &  0.69(0.03)  &  2.00(0.04)  &  1.57(0.04)  &  5.75 & --- \\ 
  BZB J0805+7534  &  J080526.63+753424.9  &  0.121  &  3.66  &  1.68  &  0.54(0.03)  &  2.02(0.04)  &  1.94(0.06)  &  1.89 & --- \\
  BZB J0809+3455  &  J080938.91+345537.3  &  0.083  &  4.07  &  ---   &  0.33(0.03)  &  1.69(0.07)  &  1.69(0.07)  &  2.40 & --- \\
  BZB J0913-2103  &  J091300.22-210321.0  &  0.198  &  12.4  &  1.94  &  0.62(0.03)  &  2.06(0.04)  &  2.57(0.07)  &  4.83 & --- \\ 
  BZB J0915+2933  &  J091552.40+293324.0  &  ?      &  6.25  &  1.87  &  0.76(0.03)  &  2.24(0.04)  &  3.31(0.09)  &  1.89 & --- \\ 
  BZB J1023-4336  &  J102356.20-433601.5  &  ?      &  13.4  &  1.82  &  0.78(0.03)  &  2.06(0.04)  &  2.16(0.06)  &  6.19 & --- \\
  BZB J1058+5628$^*$  &  J105837.73+562811.3  &  0.143  &  3.13  &  1.93  &  0.80(0.03)  &  2.28(0.03)  &  6.07(0.13)  &  0.52 & T\\  
  BZB J1117+2014  &  J111706.26+201407.5  &  0.138  &  33.6  &  1.70  &  0.61(0.03)  &  2.05(0.05)  &  1.98(0.07)  & 16.99 & C \\ 
  BZB J1120+4212  &  J112048.06+421212.6  &  0.124? &  7.81  &  1.61  &  0.70(0.03)  &  1.98(0.07)  &  1.18(0.05)  &  6.62 & --- \\ 
  BZB J1136+6737  &  J113630.10+673704.4  &  0.136  &  14.8  &  1.68  &  0.44(0.03)  &  1.87(0.06)  &  1.16(0.05)  & 12.71 & C \\ 
  BZB J1215+0732  &  J121510.98+073204.7  &  0.136  &  3.27  &  ---   &  0.42(0.04)  &  1.76(0.08)  &  1.19(0.06)  &  2.75 & --- \\
  BZB J1241-1455  &  J124149.40-145558.4  &  ?      &  8.37  &  1.98  &  0.68(0.03)  &  2.04(0.06)  &  1.51(0.06)  &  5.55 & --- \\ 
  BZB J1243+3627$^*$  &  J124312.74+362744.0  &  ?      &  10.0  &  1.70  &  0.78(0.03)  &  2.20(0.03)  &  3.97(0.09)  &  2.52 & --- \\ 
  BZB J1248+5820  &  J124818.79+582028.8  &  ?      &  3.99  &  1.95  &  0.86(0.03)  &  2.29(0.03)  &  7.59(0.15)  &  0.53 & --- \\
  BZB J1417+2543  &  J141756.67+254325.9  &  0.237  &  15.3  &  1.98  &  0.53(0.03)  &  2.02(0.05)  &  1.16(0.04)  & 13.20 & C,M \\ 
  BZB J1439+3932  &  J143917.48+393242.8  &  0.344  &  11.1  &  1.69  &  0.72(0.03)  &  2.10(0.04)  &  1.99(0.05)  &  5.57 & --- \\ 
  BZB J1443-3908  &  J144357.20-390839.9  &  0.065? &  6.56  &  1.77  &  0.73(0.03)  &  2.16(0.03)  &  4.48(0.11)  &  1.46 & --- \\
  BZB J1445-0326  &  J144506.24-032612.5  &  ?      &  3.21  &  ---   &  0.69(0.03)  &  1.86(0.08)  &  0.92(0.05)  &  3.48 & --- \\ 
  BZB J1448+3608  &  J144800.59+360831.2  &  ?      &  4.62  &  1.89  &  0.77(0.03)  &  2.13(0.04)  &  1.55(0.04)  &  2.99 & --- \\
  BZB J1501+2238  &  J150101.83+223806.3  &  0.235  &  4.02  &  1.77  &  0.83(0.03)  &  2.28(0.03)  &  6.15(0.12)  &  0.65 & --- \\
  BZB J1540+8155  &  J154015.90+815505.6  &  ?      &  5.77  &  1.48  &  0.69(0.03)  &  2.00(0.04)  &  1.40(0.04)  &  4.12 & C \\ 
  BZB J1548-2251  &  J154849.76-225102.5  &  ?      &  6.21  &  1.93  &  0.70(0.03)  &  2.11(0.05)  &  2.16(0.08)  &  2.88 & --- \\
  BZB J1725+1152  &  J172504.34+115215.5  &  ?      &  11.5  &  1.93  &  0.81(0.03)  &  2.16(0.03)  &  4.53(0.11)  &  2.54 & C \\ 
  BZB J1728+5013  &  J172818.63+501310.5  &  0.055  &  20.4  &  1.83  &  0.62(0.03)  &  2.18(0.03)  &  3.13(0.07)  &  6.52 & C,M,S \\ 
  BZB J1917-1921  &  J191744.82-192131.5  &  0.137  &  2.86  &  1.91  &  0.84(0.03)  &  2.27(0.03)  &  6.16(0.15)  &  0.46 & --- \\
  BZB J2221-5225  &  J222129.30-522527.6  &  ?      &  5.43  &  2.06  &  0.72(0.03)  &  2.17(0.05)  &  1.42(0.04)  &  3.83 & --- \\
  BZB J2323+4210  &  J232352.07+421058.6  &  0.059? &  2.69  &  1.88  &  0.77(0.03)  &  2.28(0.07)  &  1.09(0.05)  &  2.47 & --- \\
  BZB J2324-4040  &  J232444.65-404049.3  &  ?      &  14.6  &  1.81  &  0.75(0.03)  &  2.06(0.03)  &  4.77(0.11)  &  3.06 & --- \\ 
  BZB J2340+8015  &  J234054.23+801515.9  &  0.274  &  3.39  &  1.87  &  0.75(0.03)  &  2.12(0.04)  &  2.48(0.06)  &  1.37 & --- \\
\noalign{\smallskip}
\hline
\end{tabular}\\
Col. (1) \bzcat\ name. Asterisk indicates sources observed by VERITAS and not detected at TeV energies \citep[][see also Section~\ref{sec:summary}]{aliu12}.\\
Col. (2) \wse\ name.  \\
Col. (3) \bzcat\ redshift: ? = unknown, number? = uncertain.  \\
Col. (4) ROSAT X-ray flux in the 0.1-2.4 keV energy range in units of 10$^{-12}$ erg\,cm$^{-2}$\,s$^{-1}$, 
corrected for the Galactic absorption (Kalberla et al. 2005). \\
Col. (5) 2FGL $\gamma$-ray photon index $\Gamma$. \\
Cols. (6,7) IR colors from \wse. Values in parentheses are 1$\sigma$ uncertainties. \\
Col. (8) \wse\ IR flux in the 3.4-12$\mu$m energy range in units of 10$^{-12}$ erg\,cm$^{-2}$\,s$^{-1}$.  \\
Col. (9) $\Phi_{XIR}$ defined according to \S~\ref{sec:tev}. 
Col. (10) We indicate if the source was also selected by different, previous, analyses performed by 
Stecker et al. (1996; - S), Costamante \& Ghisellini (2002; - C), Tavecchio et al. (2010; - T), and Massaro et al. (2012c; - M) (see \S~\ref{sec:comparison} for more details).
\label{tab:candidates}
\end{table*}

\section{Comparison with previous selections}
\label{sec:comparison}
Several attempts to select BL Lac candidates for TeV observations have been carried out in the last decade,
with particular attention to HBLs as in our analysis.
Selection of source candidates has typically relied on the availability of source catalogs 
at lower frequencies that could reveal properties characteristic of VHE emitters. 
For instance, blazar candidates for VHE observations were typically selected 
from catalogs of hard X-ray sources \citep[e.g.,][]{stecker96,donato01} or objects that had 
a particular combination of radio, optical and X-ray energy densities \citep{costamante02}. 

In particular, Costamante \& Ghisellini (2002) proposed a selection of BL Lac candidates for TeV observations, 
not only restricted to the HBLs as in the present analysis. 
Their selection was mainly based on a fitting procedure of the broadband SEDs 
of a sample of BL Lacs compiled from literature with a homogeneous  
synchrotron self Compton model and calculating the expected TeV flux.
They conclude that TeV BL Lac candidates are primarily selected to be
bright both in X-rays and radio bands, as generally occurs for HBLs. 
Tavecchio et al. (2010) also proposed a selection of 
BL Lac candidates for TeV observations on the basis of the
$\gamma$-ray properties, such as hard $\gamma$-ray spectra, of the \fer\ sources
detected in its first three months of operation.
Recently, Massaro et al. (2011c) also outlined a criterion to select only TBCs,
mainly based on the X-ray spectral curvature and applied to the HBLs detected in the major X-ray surveys
as Einstein Slew Survey \citep[e.g.,][]{elvis92}.
This was the first criterion developed on the basis of the BL Lac spectral shape observed in a restricted energy range. 

In comparison with the analyses cited above, 8 out of 41 of the TBCs selected were also present in their lists.
We note that 2 sources appear in Stecker et al. (1996) list of candidates and 8 in that of
Costamante \& Ghisellini (2002), with 2 objects, BZB J0326+0225 and BZB J1728+5013, present in both selections.
Three of our sources appear as TeV candidates in the Tavecchio et al. (2010) selection: 
BZB J0109+1816, BZB J0136+3905 and BZB J1058+5628; the last source also in Costamante \& Ghisellini (2002).

In Massaro et al. (2011c) we also propose a X-ray based selection and  
BZB J0326+0225, BZB J1136+6737, BZB J1417+2543, and BZB J1728+5013, 
were deeply investigated and selected as TBCs on the basis of their X-ray spectral curvature.
In particular, BZB J1728+5013 is present in all the previous selections with the only exception of Tavecchio et al. (2010), 
making it the most promising TBC.

The main difference between our method and the previous selections is that
it is based on the IR rather than on the radio flux density and that was built on the basis of the peculiar IR colors of the known TBLs
(i.e., a surrogate of the IR spectral shape), an information that was not used in all the previous selections.
It is also worth noting that all $\sim$90\% BL Lacs of the \bzcat\ 
that met our criteria are also detected in the $\gamma$-rays by \fer, 
while this was not a requirement for our selection.

\section{All-sky infrared search of TBL candidates}
\label{sec:rasswise}
We extended our search of TBCs beyond
the \bzcat\ catalog by considering X-ray sources from the ROSAT bright
source catalog \citep{voges99} with a counterpart in the WISE all-sky survey \citep{wright10}
and adopting less restrictive criteria than the one previously described in \S~\ref{sec:criteria}.

We considered all the IR sources detected by \wse\ that lie within the positional uncertainty 
of an X-ray source in the ROSAT bright source catalog.
Then, we selected only IR sources with \wse\ magnitudes smaller than 13.32 mag, 12.64 mag and 10.76 mag
at 3.4$\mu$m, 4.6$\mu$m and 12$\mu$m, respectively, IR colors between 
0.23 mag $<$ [3.4]-[4.6] $<$ 0.86 mag 1.60 mag $<$ [4.6]-[12] $<$2.32 mag and
$\delta<$1 (see \S~\ref{sec:infrared} and \S~\ref{sec:criteria} for more details).
This criterion corresponds to a less restrictive selection than the one previously proposed,
because it is not based on the X-ray flux nor on the ratio $\Phi_{XIR}$.

The ROSAT bright source catalog lists 18811 X-ray sources all-sky, 
however only 189 of them met the criteria outlined above.
All of them are unique associations between the ROSAT and the \wse\ all-sky surveys.
Moreover, out of the 189 selected sources, 93 are associated to sources listed in the
\bzcat\ catalog. These were excluded from our extended TBL
candidate list to avoid redundancy in the selections.

For the remaining 96 sources, we performed a multifrequency analysis to select the most reliable TBCs.
We searched in the following major radio, IR, optical databases 
as well as in the NASA Extragalactic Database (NED)\footnote{\underline{http://ned.ipac.caltech.edu/}}
for any possible counterpart within 3\arcsec.3 to verify if additional information can confirm their BL Lac nature.
 
For the radio surveys we searched in the catalogs of the NRAO VLA Sky Survey \citep[NVSS;][]{condon98}, 
the VLA Faint Images of the Radio Sky at Twenty-Centimeters \citep[FIRST;][]{becker95,white97}, 
the Sydney University Molonglo Sky Survey \citep[SUMSS;][]{mauch03}
and the The Australia Telescope 20 GHz Survey \citep[AT20G;][]{murphy10} surveys; 
for the IR we compare our list only with the Two Micron All Sky Survey \citep[2MASS;][]{skrutskie06}
since each \wse\ source is already associated with the closest 2MASS source 
by default in the \wse\ catalog \citep[see][for more details]{cutri12}.
Then, we also searched for optical counterparts, with possible spectra available, 
in the Sloan Digital Sky Survey \citep[SDSS; e.g.][]{adelman08,paris12} and in the Six-degree-Field Galaxy Redshift Survey 
\citep[6dFGS;][]{jones04,jones09}; 
in the hard X-rays within the 2$^{nd}$ Palermo BAT catalog \citep[2PBC;][]{cusumano10} 
and if associated with \fer\ source in the 2FGL \citep{nolan12}. 
We also searched the USNO-B Catalog \citep{monet03} for the optical counterparts within 3\arcsec.3
and we report the magnitude in the R band in Table~\ref{tab:new}.
Our final list has been also compared with the recent \wse-2MASS-ROSAT selection of active galaxies proposed by Edelson \& Malkan (2012).

On the basis of our multifrequency investigation we selected 54 TBCs out of the 96 remaining sources
selected combining the ROSAT all-sky survey with the \wse\ observations.
All these new 54 TBCs are listed in Table~\ref{tab:new} together with the results of our multifrequency analysis and their salient parameters
as the IR \wse\ colors and, if present, the 2FGL association \citep[e.g.,][]{ackermann11,nolan12}.

A large fraction of our TBCs ($\sim$55\%) have a clear radio counterpart in one of the major radio surveys 
as occurs for all the BL Lacs that belong to the \wse\ Gamma-ray strip.
{ In particular, 21 out of 54 sources have a radio counterpart in the NVSS, 4 in the SUMSS, one in both radio surveys while only 2 objects
have a correspondence in the FIRST. All our 54 TBCs are detected in the 2MASS catalog and they are also detected by \wse\ at 22$\mu$m
with two exceptions. Optical spectra are available in literature for 5 TBCs listed in Table~\ref{tab:new}
all classified as BL Lac objects, thus no optical spectroscopic observations are necessary to confirm their nature.
Then, 21 out of 54 TBCs have been observed by the 6dFGS and 4 by the SDSS.
At high energies only one source has been detected by \swf-BAT hard X-ray survey while 14 out of 54 objects 
have been associated or lie within the positional uncertainty region of a \fer\ source listed in the 2FGL.}
It is worth noting that 14 out of 54 TBCs are associated with \fer\ sources as listed in the 2FGL catalog and in the 2LAC catalogs,
being classified as AGNs of uncertain type \citep[][]{nolan12,ackermann11}.
In particular, 1RXS J083158.1-180828, 1RXS J130421.2-435308 and 1RXS J204745.9-024609 are the only three sources that belong to 
our final list of TBCs selected combining \wse\ and ROSAT all-sky catalogs that were also selected as active galaxies 
of uncertain nature by Edelson \& Malkan (2012).

Moreover, 16 X-ray sources out of 54 TBCs were previously unidentified in the ROSAT all-sky catalog \citep{voges99}
(i.e., without a counterpart assigned at lower energies).
The existence of an IR \wse\ counterpart in the \wse\ catalog, with similar colors than those of $\gamma$-ray BL Lacs
suggests that these could be potential BL Lac candidates.
We note that none of these 54 TBCs is listed in the Sedentary Survey of extreme HBLs \citep{giommi05},
thus highlighting that our method is successful to select BL Lacs without including any criteria based on the radio observations.

Forty-two out of 189 IR-X-ray sources we selected have multiwavelength archival observations that indicate they are not BL Lacs.
This suggest that a contamination of $\sim$22\% of non-BL Lac objects could be present in our selected sample.
We will be able to make a more accurate estimate of the contamination once all the optical spectroscopic information will be available
for our sample.

Finally, we note that all the TBCs selected from the ROSAT bright source catalog with a counterpart in the WISE all-sky survey
lie within 3$\sigma$ level of confidence of the regression line (see Section~\ref{sec:infrared}),
with the only exceptions of 5 sources, namely:
1RXS J072812.1+671821, 1RXS J132908.3+295018, 1RXS J180219.5-245157, 1RXS J183821.0-602519 and 1RXS J193320.3+072616.

\section{Summary and conclusions}
\label{sec:summary}
Previous studies based on the \wse\ all-sky survey have
revealed that the IR spectral shape of high frequency peaked BL Lacs detected at TeV energies (TBLs)
can be successfully used to associate $\gamma$-ray BL Lacs \citep[e.g.,][]{paper1,paper4,paper6}. 
In this paper, we extended the same technique to search for
high frequency peaked BL Lacs that could be candidates for future TeV observations (i.e., TBCs),
by selecting sources with similar IR and X-ray properties of the known TBLs.

Known TBLs populate a subregion of the \wse\ Gamma-ray
Strip \citep{paper3,paper6}, defined in the [3.4]-[4.6]-[12] $\mu$m IR color-color space. 
Then, on the basis of their IR and the X-ray emission,
we identify 41 TBCs among the BL Lacs listed in the \bzcat\ catalog \citep{massaro09,massaro11}.

A comparison between our list of TBCs, chosen out of the \bzcat\ catalog, with previous selections 
\citep{stecker96,costamante02,tavecchio10,massaro11c} finds good agreement (see \S~\ref{sec:comparison}).
Our new criteria, mainly based on the IR colors,
a surrogate of the spectral shape of the low energy component for the BL Lac objects, 
is not based on radio or $\gamma$-ray data.
Moreover, our IR selection was built only in the 2-dimensional [3.4]-[4.6]-[12]$\mu$m 
color-color diagram, { while all our previous selections} of $\gamma$-ray blazar candidates, 
mostly developed to associate unidentified gamma-ray sources \citep{paper4,paper6},
required the IR detection in all four \wse\ bands.
All the BL Lacs of the \bzcat\ that met our criteria are also detected in the $\gamma$-ray band by \fer, 
while this was not a requirement for the selection discussed in this paper.

We note that VERITAS observations have been performed for 3 of our TBCs selected within the \bzcat:
BZB J0136+3905, BZB J1058+5628 and BZB J1243+3627 \citep{aliu12}.
However, as discussed in Aliu et al. (2012), both BZB J0136+3905 and BZB J1243+3627
do not have a redshift estimate, indicating that their non TeV-detection could be due 
the absorption of high energy photons by the extragalactic background light \citep{franceschini08}.
On the other hand, BZB J1058+5628 was found variable in the $\gamma$-rays by \fer\ during the same period of the VERITAS observations.
Thus, the non-detection at TeV energies of these three sources 
does not affect our selection that can only be verified with additional TeV observations.

We conducted an extended search based on less restrictive criteria based on the combination of the \wse\ and ROSAT observations
to search for new TBCs with IR properties similar to those of the TBLs in the X-ray sky.
We found additional 54 sources that could be considered TBCs with pending confirmation of their
BL Lac nature with follow-up optical spectroscopy.
We also note that 16 TBCs out of 54 X ray sources were previously unidentified in the ROSAT bright source 
catalog \citep{voges99}; then we provide the first association with a low energy counterpart
correspondent to our TBCs selected on the basis of the IR \wse\ colors.  
We note that only 21 out of a total of 95 TBCs 
(i.e., 41 selected in the \bzcat\ and 54 in the all-sky search) have a reliable redshift determination.
The TBC with the highest redshift is BZB J0316-2607 at z=0.443 closer
than the most distant TeV source: 3C\,279 at z=0.5362 \citep[e.g.,][]{errando08}.

Our investigation provides new targets to plan observations 
with ground based Cherenkov telescopes such as HESS, MAGIC and VERITAS or in the near future with CTA. 

\begin{table*}
\tiny
\caption{TBCs all-sky selected from the ROSAT - \wse\ all-sky surveys (00 -- 24 HH)}
\begin{tabular}{|lllcccclc|}
\hline
  ROSAT  &  \wse\  &  other & [3.4]-[4.6] & [4.6]-[12] & [12]-[22] & R & notes & z \\
  name  &  name   &  name  &     mag     &    mag     &   mag     & mag &              &   \\
\hline
\noalign{\smallskip}
J001541.3+555141  &  J001540.13+555144.7  &  NVSS J001540+555144  &  0.59(0.03) &  1.93(0.05) &  2.14(0.15) & 16.07 & N,M          &  ?\\
J002159.2-514028  &  J002200.08-514024.2  &  SUMSS J002159-514026 &  0.81(0.03) &  2.23(0.04) &  1.87(0.14) & 15.94 & S,M,6,v,f    &  ?\\
J002922.4+505159  &  J002921.68+505159.0  &                       &  0.86(0.03) &  2.22(0.05) &  2.12(0.15) & 16.61 & M,UNID       &  ?\\
J005447.2-245532  &  J005446.74-245529.0  &  NVSS J005446-245529  &  0.74(0.04) &  1.95(0.08) &  $<$1.82    & 17.08 & N,M,6,f,BL   &  ?\\
J010325.9+533721  &  J010325.95+533713.3  &  NVSS J010326+533712  &  0.45(0.04) &  1.91(0.06) &  1.74(0.26) & 14.94 & N,M,f        &  ?\\
J013445.2-043017  &  J013445.62-043012.9  &  6dF J0134455-043013  &  0.74(0.03) &  2.21(0.03) &  2.19(0.04) & 13.41 & N,M,6,v      &  ?\\
J014100.4-675332  &  J014100.45-675327.2  &  6dF J0141003-675328  &  0.60(0.03) &  1.85(0.05) &  0.77(0.51) & 15.83 & M,6,v        &  ?\\
J021652.4-663644  &  J021650.85-663642.5  &  SUMSS J021650-663643 &  0.73(0.03) &  2.12(0.05) &  2.18(0.13) & 17.14 & S,M,6,f      &  ?\\
J024215.2+053037  &  J024214.63+053036.0  &  NVSS J024214+053042  &  0.77(0.03) &  2.32(0.03) &  1.82(0.05) & 12.03 & N,M,v,B      &  ?\\
J032220.5-305929  &  J032220.09-305933.9  &  6dF J0322201-305934  &  0.79(0.03) &  2.17(0.04) &  1.97(0.15) & 16.00 & M,6,v        &  ?\\
J033118.2-615532  &  J033118.46-615528.8  &  6dF J0331185-615529  &  0.66(0.03) &  2.14(0.04) &  1.84(0.17) & 16.20 & M,6          &  ?\\
J033913.4-173553  &  J033913.70-173600.6  &  NVSS J033913-173600  &  0.32(0.03) &  1.72(0.04) &  2.00(0.16) & 10.98 & N,A,M,6,f    &  0.0656?\\
J034203.8-211428  &  J034203.71-211439.3  &  NVSS J034203-211449  &  0.68(0.03) &  2.24(0.03) &  1.82(0.02) &  9.27 & N,M,6        &  ?\\
J043917.9+224802  &  J043917.42+224753.3  &                       &  0.46(0.03) &  1.80(0.03) &  1.49(0.03) & 13.03 & M            &  ?\\
J045142.3-034834  &  J045141.51-034833.6  &  NVSS J045141-034834  &  0.55(0.03) &  2.00(0.03) &  1.96(0.04) &  8.25 & N,M,6,B      &  ?\\
J051952.0-512347  &  J051952.79-512338.0  &                       &  0.73(0.03) &  2.30(0.03) &  2.13(0.06) & 14.72 & M            &  ?\\
J062040.0+264339  &  J062040.05+264331.9  &  NVSS J062040+264331  &  0.51(0.03) &  1.88(0.08) &  2.23(0.25) & 15.80 & N,M          &  ?\\
J062221.4-260537  &  J062222.06-260544.6  &  NVSS J062222-260544  &  0.82(0.03) &  2.29(0.04) &  2.10(0.11) & 16.94 & N,M,6,f,BL   &  ?\\
J063923.6+010231  &  J063923.53+010231.2  &                       &  0.79(0.04) &  2.28(0.06) &  2.55(0.12) & 16.01 & M,UNID       &  ?\\
J064007.4-125316  &  J064007.19-125315.0  &  NVSS J064007-125315  &  0.46(0.03) &  1.74(0.04) &  1.95(0.15) & 13.69 & N,A,M,6      &  ?\\
J065610.6+460538  &  J065609.67+460541.5  &                       &  0.80(0.03) &  2.29(0.04) &  2.22(0.10) & 15.17 & M,UNID       &  ?\\
J070912.3-152708  &  J070912.51-152703.6  &  NVSS J070912-152701  &  0.48(0.03) &  1.63(0.05) &  2.35(0.14) & 15.54 & N,M          &  ?\\
J072259.5-073131  &  J072259.68-073135.0  &  NVSS J072259-073135  &  0.75(0.04) &  1.98(0.06) &  2.10(0.24) & 16.77 & N,M          &  ?\\
J072812.1+671821  &  J072812.88+671814.7  &                       &  0.78(0.03) &  1.97(0.06) &  2.04(0.21) & 16.56 & M,UNID       &  ?\\ 
J073048.2-660212  &  J073049.52-660218.9  &                       &  0.47(0.03) &  1.88(0.04) &  2.23(0.09) & 14.95 & M,UNID       &  ?\\
J073143.9-470009  &  J073144.11-470008.4  &                       &  0.54(0.08) &  1.92(0.06) &  2.06(0.08) & ---   & M,UNID       &  ?\\
J082705.9-070841  &  J082706.17-070845.9  &  NVSS J082706-070846  &  0.64(0.03) &  1.97(0.04) &  1.64(0.17) & 14.75 & N,M,6,BL     &  0.12?\\
J083158.1-180828  &  J083158.37-180835.2  &  6dF J0831584-180835  &  0.82(0.03) &  2.24(0.04) &  2.29(0.08) & 15.74 & M,6          &  ?\\ 
J094709.2-254056  &  J094709.52-254100.0  &  NVSS J094709-254100  &  0.73(0.04) &  2.17(0.06) &  2.06(0.23) & 16.71 & N,M,6,f      &  ?\\
J130421.2-435308  &  J130421.01-435310.2  &  SUMSS J130420-435308 &  0.86(0.03) &  2.27(0.03) &  1.93(0.05) & 16.10 & S,M,f,v      &  ?\\ 
J130737.8-425940  &  J130737.98-425938.9  &  SUMSS J130737-425940 &  0.74(0.03) &  2.06(0.03) &  1.90(0.08) & 15.58 & S,M,6,f,v    &  ?\\
J132452.1+213559  &  J132451.92+213548.7  &SDSSJ132451.91+213548.8&  0.67(0.03) &  2.17(0.04) &  2.04(0.13) & 14.67 & F,M,s        &  ?\\
J132908.3+295018  &  J132908.84+295024.2  &SDSSJ132908.83+295024.2&  0.61(0.03) &  2.21(0.06) &  1.94(0.21) & 14.75 & M,s          &  ?\\ 
J134751.3+283639  &  J134751.55+283631.5  &SDSSJ134751.52+283632.3&  0.36(0.03) &  1.73(0.06) &  2.65(0.15) & 14.47 & F,M,s        &  ?\\
J140906.7-451714  &  J140907.20-451715.8  &  6dF J1409074-451716  &  0.56(0.03) &  1.84(0.04) &  1.81(0.15) & 14.25 & M,6,v        &  ?\\
J150554.3-694935  &  J150555.68-694932.6  &                       &  0.85(0.03) &  2.16(0.05) &  2.33(0.10) & 17.55 & M,UNID       &  ?\\
J153548.6-295904  &  J153548.53-295855.5  &  6dF J1535486-295854  &  0.34(0.03) &  1.73(0.03) &  1.98(0.04) & 13.38 & M,6          &  ?\\
J154513.6-341733  &  J154512.84-341730.6  &                       &  0.75(0.06) &  2.08(0.03) &  2.04(0.02) &  9.63 & M,UNID       &  ?\\
J170034.7-273807  &  J170034.97-273804.4  &                       &  0.45(0.05) &  1.65(0.03) &  2.14(0.03) & 10.40 & M,UNID       &  ?\\
J180219.5-245157  &  J180219.45-245154.3  &                       &  0.51(0.03) &  1.61(0.02) &  2.29(0.02) & ---   & M,v,UNID     &  ?\\ 
J180925.6+204130  &  J180925.43+204131.2  &  NVSS J180925+204131  &  0.80(0.04) &  2.12(0.06) &  1.52(0.33) & 16.49 & N,M,f        &  ?\\
J182022.7-101104  &  J182022.75-101113.4  &                       &  0.33(0.04) &  1.63(0.03) &  2.12(0.02) & 10.12 & M,UNID       &  ?\\
J182339.2-345412  &  J182338.59-345412.0  &  NVSS J182338-345412  &  0.70(0.04) &  2.07(0.04) &  1.92(0.13) & ---   & N,A,M,f      &  ?\\
J183821.0-602519  &  J183820.64-602522.4  &                       &  0.33(0.03) &  1.83(0.07) &  2.26(0.19) & 13.11 & M,UNID       &  ?\\  
J184121.8+290932  &  J184121.73+290940.9  &  NVSS J184121+290945  &  0.61(0.04) &  2.09(0.06) &  1.92(0.26) & 16.66 & N,M          &  ?\\
J192503.1+504315  &  J192502.18+504313.8  &                       &  0.69(0.03) &  2.20(0.03) &  2.35(0.04) & 14.68 & M,UNID       &  ?\\
J192649.5+615445  &  J192649.89+615442.4  &  NVSS J192649+615441  &  0.79(0.03) &  2.18(0.04) &  1.88(0.13) & 17.20 & N,M,f        &  ?\\
J193320.3+072616  &  J193320.30+072621.9  &  NVSS J193320+072619  &  0.85(0.05) &  2.07(0.08) &  1.90(0.34) & 16.60 & N,M,UNID     &  ?\\ 
J195020.5+331419  &  J195019.72+331416.2  &                       &  0.85(0.03) &  2.17(0.03) &  1.75(0.12) & 15.82 & M,UNID       &  ?\\
J195815.6-301119  &  J195814.91-301111.5  &  NVSS J195814-301112  &  0.42(0.03) &  1.86(0.07) &  2.12(0.25) & 13.97 & N,S,M,s,6,f,BL&  ?\\
J204149.8-373346  &  J204150.23-373339.8  &  6dF J2041502-373340  &  0.31(0.03) &  1.63(0.08) &  $<$2.25    & 13.62 & M,6,BL       &  0.0986\\
J204745.9-024609  &  J204745.80-024604.1  &  NVSS J204745-024605  &  0.85(0.03) &  2.32(0.04) &  1.98(0.09) & 15.01 & N,A,M,6      &  ?\\ 
J224427.7+440135  &  J224427.24+440137.4  &                       &  0.72(0.03) &  2.32(0.03) &  2.12(0.04) & 13.43 & M,UNID       &  ?\\
J224753.3+441321  &  J224753.19+441315.6  &  NVSS J224753+441317  &  0.69(0.03) &  2.13(0.06) &  1.94(0.22) & 16.75 & N,M,f        &  ?\\
\noalign{\smallskip}
\hline
\end{tabular}\\
Col. (1) ROSAT name. \\
Col. (2) \wse\ name.  \\
Col. (3) Other name if present in literature and in the following order: NVSS, SDSS, AT20G, NED.  \\
Cols. (4,5,6) IR colors from \wse. Values in parentheses are 1$\sigma$ uncertainties. \\
Col. (7) Notes: N = NVSS, F = FIRST, S = SUMSS, A=AT20G, M = 2MASS, s = SDSS dr9, 6 = 6dFGS, x = \xmm\ or \chn, X = ROSAT;  B=\swf-BAT; f=\fer; 
BL = BL Lac (optical spectra available in Jones et al. 2009); v = variability in \wse\ (var\_flag $>$ 5 in at least one band); UNID=ROSAT unidentified X-ray source.  \\
Col. (10) Redshift: ? = unknown, number? = uncertain. 
\label{tab:new}
\end{table*}

\acknowledgements
We thank the anonymous referee for useful comments that improved the presentation of our work.
We are grateful to R. Mukherjee for her valuable comments and suggestions that improved the manuscript
as well as  to S. Digel, P. Giommi, D. Harris and J. Grindlay for their helpful discussions.
The work is supported by the NASA grants NNX12AO97G.
R. D'Abrusco gratefully acknowledges the financial 
support of the US Virtual Astronomical Observatory, which is sponsored by the
National Science Foundation and the National Aeronautics and Space Administration.
The work by G. Tosti is supported by the ASI/INAF contract I/005/12/0.
M. Errando acknowledges support from the NASA grant NNX12AJ30G.
TOPCAT\footnote{\underline{http://www.star.bris.ac.uk/$\sim$mbt/topcat/}} 
\citep{taylor2005} was used extensively in this work 
for the preparation and manipulation of the tabular data and the images.
Part of this work is based on archival data, software or on-line services provided by the ASI Science Data Center.
This research has made use of data obtained from the High Energy Astrophysics Science Archive
Research Center (HEASARC) provided by NASA's Goddard
Space Flight Center; the SIMBAD database operated at CDS,
Strasbourg, France; the NASA/IPAC Extragalactic Database
(NED) operated by the Jet Propulsion Laboratory, California
Institute of Technology, under contract with the National Aeronautics and Space Administration.
Part of this work is based on the NVSS (NRAO VLA Sky Survey);
The National Radio Astronomy Observatory is operated by Associated Universities,
Inc., under contract with the National Science Foundation. 
This publication makes use of data products from the Two Micron All Sky Survey, which is a joint project of the University of 
Massachusetts and the Infrared Processing and Analysis Center/California Institute of Technology, funded by the National Aeronautics 
and Space Administration and the National Science Foundation.
This publication makes use of data products from the Wide-field Infrared Survey Explorer, 
which is a joint project of the University of California, Los Angeles, and 
the Jet Propulsion Laboratory/California Institute of Technology, 
funded by the National Aeronautics and Space Administration.

{}

\begin{figure}[!htp]
\begin{center}
\includegraphics[height=8.5cm,width=6.5cm,angle=-90]{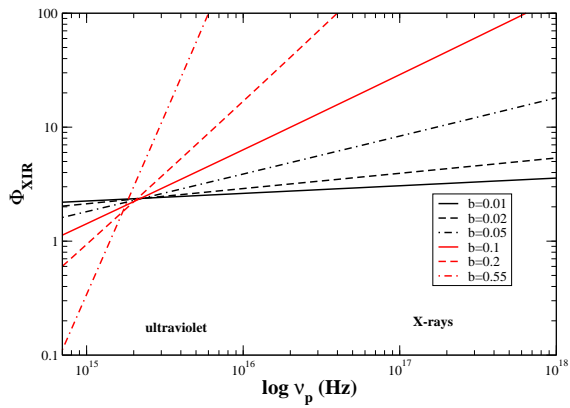}
\caption{The values of $\Phi_{XIR}$ as function of the peak frequency $\nu_p$
of the log-parabolic function (see Eq. 1) assumed as simple representation of the low energy component of the BL Lac SED.
The different lines correspond to different values of the spectral curvature $b$, typical of BL Lac objects
\citep[e.g.,][]{massaro08a,massaro08b}.}
\label{fig:phi}
\end{center}
\end{figure}

\appendix
To further justify the classification scheme proposed in Section~\ref{sec:phi} based on $\Phi_{XIR}$, 
we assumed a broadband description of the BL Lac spectra, from the IR to the X-rays, 
in the form of a log-parabola \citep[e.g.,][]{howard65,landau86}, expressed as:
\begin{equation}
S_{\nu} = \frac{S_p}{\nu} \cdot \left(\frac{\nu}{\nu_p}\right)^{-b~log(\nu/\nu_{p})} \, erg\,cm^{-2}\,s^{-1}\,Hz^{-1}
\end{equation}
where $\nu_p$ is the SED peak frequency, $S_p$ the SED peak flux at $\nu_p$, and $b$ the spectral curvature
\citep[see][for more details]{massaro04,tramacere07}.
We computed the ratio $\Phi_{XIR}$ as function of the peak frequency $\nu_p$ for different values of $b$ as shown
in Figure~\ref{fig:phi}. Thus for values of $\nu_p$ larger than 10$^{15}$Hz, as generally seen for HBLs, 
$\Phi_{XIR}$ is systematically larger than 0.1 (see Figure~\ref{fig:phi}).
Values of spectral curvature used in Figure~\ref{fig:phi} are those typically observed in BL Lac objects
\citep{massaro08a,massaro08b,massaro11}.


\begin{thebibliography}{}
\bibitem[Abdo et al. 2009]{abdo09} Abdo, A. A., et al. 2009, Astroparticle Physics, 32, 193 
\bibitem[Abdo et al. 2010]{abdo10} Abdo, A. A. et al. 2010 ApJS 188 405 
\bibitem[Ackermann et al. 2011]{ackermann11} Ackermann, M. et al. 2011 ApJ, 743, 171 
\bibitem[Ackermann et al. 2012]{ackermann12} Ackermann, M. et al. 2012 ApJ, 753, 83 
\bibitem[Adelman et al. 2008]{adelman08} Adelman-McCarthy, J., Agueros, M.A., Allam, S.S., et al. 2008, ApJS, 175, 297 
\bibitem[Aliu et al. 2012]{aliu12} Ali, E. et al. 2012 ApJ, 759, 102
\bibitem[Becker et al. 1995]{becker95} Becker, R. H., White, R. L., Helfand, D. J.1995 ApJ, 450, 559
\bibitem[Blandford \& Rees 1978]{blandford78} Blandford, R. D., Rees, M. J., 1978b PhyS, 17, 265
\bibitem[Blandford \& K$\ddot{o}$nigl 1979]{blandford79} Blandford, R. D. \& K$\ddot{o}$nigl, A.1979 ApJ, 232, 34
\bibitem[Condon et al. 1998]{condon98} Condon, J. J., Cotton, W. D., Greisen, E. W., Yin, Q. F., Perley, R. A., Taylor, G. B., \& Broderick, J. J. 1998, AJ, 115, 1693 
\bibitem[Costamante \& Ghisellini 2002]{costamante02} Costamante, L. \& Ghisellini, G. 2002 A\&A, 384, 56
\bibitem[Cusumano et al. 2010]{cusumano10} Cusumano, G. et al. 2010 A\&A, 524A, 64
\bibitem[Cutri et al. 2012]{cutri12} Cutri et al. 2012 wise.rept, 1C
\bibitem[D'Abrusco et al. 2012]{paper2} D'Abrusco, R., Massaro, F., Ajello, M., Grindlay, J. E., Smith, Howard A. \& Tosti, G. 2012 ApJ, 748, 68 
\bibitem[D'Abrusco et al. 2013]{paper6} D'Abrusco, R., Massaro, F., Paggi, A., Masetti, N., Giroletti, M., Tosti, G. et al. 2013 ApJS submitted 
\bibitem[Donato et al. 2001]{donato01} Donato, D., Ghisellini, G., Tagliaferri, G. \& Fossati, G., 2001, A\&A, 375, 739
\bibitem[Edelson \& Malkan 2012]{edelson12} Edelson, R. \& Malkan, M. 2012 ApJ, 751, 52
\bibitem[Elvis et al.1992]{elvis92} Elvis, M., Plummer, D., Schachter, J., Fabbiano, G. 1992 ApJS, 80, 257
\bibitem[Errando et al. 2008]{errando08} Errando, M., Bock, R., Kranich, D., Lorenz, E., Majumdar, P., Mariotti, M., Mazin, D., Prandini, E. et al. 2008 AIPC, 1085, 423
\bibitem[Franceschini et al. 2008]{franceschini08} Franceschini, A., Rodighiero, G., Vaccari, M.	2008 A\&A, 487, 837
\bibitem[Giommi et al. 2005]{giommi05} Giommi, P. et al. 2005, A\&A, 434, 385
\bibitem[Giommi et al. 2007]{giommi07} Giommi, P. et al. 2007 A\&A, 468, 571
\bibitem[Giommi et al. 2009]{giommi09} Giommi, P., Colafrancesco, S., Padovani, P., Gasparrini, D., Cavazzuti, E., Cutini, S. 2009, A\&A, 508, 107
\bibitem[Giommi et al. 2012a]{giommi12a} Giommi, P. et al. 2012a A\&A, 541A, 160
\bibitem[Giommi et al. 2012b]{giommi12b} Giommi, P., Padovani, P., Polenta, G., Turriziani, S., D'Elia, V., Piranomonte, S. 2012b MNRAS, 420, 2899
\bibitem[Gonz\'alez-Nuevo et al. 2010]{gonzales10} Gonz\'alez-Nuevo, J. et al. 2010 A\&A, 518, L38
\bibitem[Hartman et al. 1999]{hartman99} Hartman, R.C. et al., 1999 ApJS 123 
\bibitem[Howard et al. 1965]{howard65} Howard, W. E. III, Dennis, T. R., Maran, S. P.; Aller, H. D. 1965 ApJS, 10, 331
\bibitem[Kalberla et al. 2005]{kalberla05} Kalberla, P.M.W., Burton, W.B., Hartmann, D., 2005, A\&A, 440, 775
\bibitem[Impey et al. 1988]{impey88} Impey, C. D. \& Neugebauer, G. 1988 AJ, 95, 307
\bibitem[Jones et al. 2004]{jones04} Jones, H. D. et al. 2004 MNRAS, 355, 747
\bibitem[Jones et al. 2009]{jones09} Jones, H. D. et al. 2009 MNRAS, 399, 683 
\bibitem[Landau et al. 1986]{landau86} Landau, R., Golish, B., Jones, T. J., et al. 1986, ApJ, 308, L78
\bibitem[Laurent-Muehleisen et al. 1999]{laurent99} Laurent-Muehleisen, S. A., Kollgaard, R. I., Feigelson, E. D., Brinkmann, W., Siebert, J. 1999 ApJ, 525, 127
\bibitem[Maselli et al. 2010a]{maselli10a} Maselli, A., Massaro, E., Nesci, R., Sclavi, S., Rossi, C., Giommi, P. 2010a A\&A, 512A, 74
\bibitem[Maselli et al. 2010b]{maselli10b} Maselli, A., Cusumano, G., Massaro, E., La Parola, V., Segreto, A., Sbarufatti, B. 2010b A\&A, 520A, 47
\bibitem[Massaro et al. 2004]{massaro04} Massaro, E., Perri, M., Giommi, P., et al. 2004, A\&A, 422, 103
\bibitem[Massaro et al. 2008a]{massaro08a} Massaro, F. et al. 2008a A\&A, 489, 1047
\bibitem[Massaro et al. 2008b]{massaro08b} Massaro, F., Tramacere, A., Cavaliere, A., Perri, M., Giommi, P. 2008b A\&A, 478, 395
\bibitem[Massaro et al. 2009]{massaro09} Massaro, E., Giommi, P., Leto, C., Marchegiani, P., Maselli, A., Perri, M., Piranomonte, S., Sclavi, S. 2009 A\&A, 495, 691 
\bibitem[Massaro et al. 2010]{massaro10} Massaro, E., Giommi, P., Leto, C., Marchegiani, P., Maselli, A., Perri, M., Piranomonte, S., Sclavi, S. 2010 {http://arxiv.org/abs/1006.0922}  
\bibitem[Massaro et al. 2011a]{paper1} Massaro, F., D'Abrusco, R., Ajello, M., Grindlay, J. E. \& Smith, H. A. 2011 ApJ, 740L, 48 
\bibitem[Massaro et al. 2011b]{massaro11} Massaro, E., Giommi, P., Leto, C., Marchegiani, P., Maselli, A., Perri, M., Piranomonte, S., 2011 ``Multifrequency Catalogue of Blazars (3rd Edition)", 2011a ARACNE Editrice, Rome, Italy  
\bibitem[Massaro et al. 2011c]{massaro11c} Massaro, F., Paggi, A., Elvis, M., Cavaliere, A. 2011 ApJ, 739, 73
\bibitem[Massaro et al. 2012a]{paper3} Massaro, F., D'Abrusco, R., Tosti, G., Ajello, M., Gasparrini, D., Grindlay, J. E. \& Smith, Howard A. 2012b ApJ, 750, 138 
\bibitem[Massaro et al. 2012b]{paper4} Massaro, F., D'Abrusco, R., Tosti, G., Ajello, M., Paggi, A., Gasparrini, 2012c ApJ, 752, 61  
\bibitem[Massaro et al. 2013]{paper} Massaro, F. et al. 2013 ApJS in preparation 
\bibitem[Mauch et al. 2003]{mauch03} Mauch, T., Murphy, T., Buttery, H. J., Curran, J., Hunstead, R. W., Piestrzynski, B., Robertson, J. G., Sadler, E. M. 2003 MNRAS, 342, 1117
\bibitem[Monet et al. 2003]{monet03} Monet, D. G. et al. 2003 AJ, 125, 984 
\bibitem[Mukherjee et al. 1997]{mukherjee97} Mukherjee, R. et al., 1997 ApJ, 490, 116
\bibitem[Murphy et al. 2010]{murphy10} Murphy, T. et al. 2010 MNRAS, 402, 2403 
\bibitem[Nolan et al. 2012]{nolan12} Nolan et al. 2012 ApJS, 199, 31 
\bibitem[Paris et al. 2012]{paris12} Paris, I. et al. 2012 A\&A, 548A, 66
\bibitem[Schneider et al. 2007]{schneider07} Schneider et al. 2007, AJ, 134, 102
\bibitem[Skrutskie et al. 2006]{skrutskie06} Skrutskie, M. F. et al. 2006, AJ, 131, 1163 
\bibitem[Stickel et al. 1991]{stickel91} Stickel, M., Padovani, P., Urry, C. M., Fried, J. W., Kuehr, H. 1991 ApJ, 374, 431
\bibitem[Stoke et al. 1991]{stoke91} Stocke et al. 1991, ApJS, 76, 813
\bibitem[Su \& Finkbeiner 2012]{su12} Su, M. \& Finkbeiner, D. P. 2012 ApJ submitted http://arxiv.org/abs/1207.7060v1
\bibitem[Stecker et al. 1996]{stecker96} Stecker, F. W.; de Jager, O. C.; Salamon, M. H. 1996 ApJ, 473L, 75
\bibitem[Urry \& Padovani 1995]{urry95} Urry, C. M., \& Padovani, P. 1995, PASP, 107, 803
\bibitem[Tavecchio et al. 2010]{tavecchio10} Tavecchio, F., Ghisellini, G., Ghirlanda, G., Foschini, L., Maraschi, L. 2010 MNRAS, 401,1570
\bibitem[Taylor 2005]{taylor2005} Taylor, M. B. 2005, ASP Conf. Ser., 347, 29 
\bibitem[Thompson 2008 et al. 2012]{thompson08}2008 RPPh, 71k6901
\bibitem[Tramacere et al. 2007]{tramacere07} Tramacere, A., Massaro, F., Cavaliere, A., 2007, A\&A, 466, 521 
\bibitem[Voges et al. 1999]{voges99} Voges, W. et al. 1999 A\&A, 349, 389
\bibitem[White et al. 1997]{white97} White, R. L., Becker, R. H. Helfand, D. J., Gregg, M. D. et al. 1997 ApJ, 475, 479 
\bibitem[Wright et al. 2010]{wright10} Wright, E. L., et al. 2010 AJ, 140, 1868
\bibitem[Zechlin et al. 2012]{zechlin12} Zechlin, H.-S., Fernandes, M. V., Elsasser, D., Horns, D. 2012 A\&A, 538A, 93
\end{thebibliography}
\end{document}